\documentclass[aps,twocolumn,showpacs,tightenlines,superscriptaddress]{revtex4}

\usepackage{amsmath}
\usepackage{amsfonts}
\usepackage{amssymb}
\usepackage{graphicx}

\begin{document}

\title{Multipartite Bell's inequalities involving many measurement settings}
\author{{\v C}aslav Brukner}
\affiliation{Institut f\"ur Experimentalphysik, Universit\"at
Wien, Boltzmanngasse 5, A--1090 Wien, Austria}
\author{Wies{\l}aw Laskowski}
\affiliation{Instytut Fizyki Teoretycznej i Astrofizyki
Uniwersytet Gda\'nski, PL-80-952 Gda\'nsk, Poland}
\author{Tomasz Paterek}
\affiliation{Instytut Fizyki Teoretycznej i Astrofizyki
Uniwersytet Gda\'nski, PL-80-952 Gda\'nsk, Poland}
\author{Marek {\. Z}ukowski}
\affiliation{Instytut Fizyki Teoretycznej i Astrofizyki
Uniwersytet Gda\'nski, PL-80-952 Gda\'nsk, Poland}

\date{\today}

\begin{abstract}
We present a prescription for obtaining Bell's inequalities
for $N\!>\!2$ observers involving {\it more than two} alternative
measurement settings. We give examples of some families of such inequalities. 
The inequalities are violated by certain classes of states 
for which all standard Bell's inequalities with two measurement settings per
observer are satisfied.
\end{abstract}


\maketitle

Which quantum states do not allow a local realistic
description \footnote{Realism supposes that measurement results are predetermined by the
properties the particles carry prior to and independent of observations.
Locality supposes that these results are independent of any
action at space-like separations.}? Despite the considerable research efforts, this
question remains open. One of the reasons for this is that
our present tools to test local realism are not optimal. Most of
Bell's inequalities are for the case in which only {\it two}
measurement settings can be chosen by each observer. One can call
such inequalities "standard" ones. The usual
Bell inequalities for bipartite two-dimensional systems
\cite{BELL,CH} including the Clauser-Horne-Shimony-Holt (CHSH)
inequality \cite{CHSH} and inequalities for bipartite
higher-dimensional systems \cite{KAZ} are standard ones.
Similarly, multipartite Bell's inequalities like those of Mermin
and Klyshko \cite{MERMIN} and the recently found set of
multipartite Bell inequalities for correlation functions
\cite{WZB,WW} also belong to the class.

One can expect that the full set of Bell inequalities for the case
when the observers can choose between more than two observables
should give a {\it more stringent constraints} on local realistic
description of quantum predictions than the standard
inequalities. 

The derivation of such constraints is important for several reasons:
First, the new inequalities can shed new light on the subtle
relation between non-separability (impossibility to
decompose quantum state of a composite system as a convex sum of product
states) and violation of local realism. They may {\it
extend the class of non-separable states which cannot be described by local
realistic models}. There are non-separable mixed
states that admit local realistic description \cite{WERNER}, but
the recent results show increasing subtlety in
this relation, as the number of systems grows. Some
multi-particle {\it pure entangled} states satisfy all
standard Bell correlation function inequalities \cite{ZBLW}. For example, the
generalized GHZ states
\begin{equation}
|\psi\rangle = \cos{\alpha |111\rangle} +
\sin{\alpha|000\rangle}, \label{state}
\end{equation}
satisfy all standard inequalities in the range $\alpha \in
[0,\pi/12]$ \cite{SCARANI,ZBLW}.
Second, the violation of Bell's inequalities is an important
primitive for building quantum information protocols that
decrease the communication complexity \cite{BZJZ,MASSAR} and is a
criterion for the efficient extraction of secure key in quantum
key distribution protocols \cite{SCARANI1}. On the basis of
Bell's inequalities involving many measurement settings one can
expect to build new quantum communication complexity protocols and
strengthen the criteria for secure quantum key distribution.

Thus far we still lack a general and efficient method for deriving
Bell inequalities involving more than two measurement settings
per observer. One possible approach is to compute 
Bell's inequalities that define the facets 
of the full correlation polytope
\cite{PITOWSKYSVOZIL}. Using this method a bipartite Bell
inequality was found with 3 measurement settings per party, 
which can be violated even when the CHSH inequality is satisfied
\cite{COLLINS}. Yet this approach is limited by its connection to
the computationally hard NP-problems \cite{PITOWSKY} and thus
applicable only for small numbers of parties, measurement
settings and outcomes \footnote{Because of the explosive growth of the number 
of inequalities, with growing number of particles, even the task 
of printing of the inequalities is a NP-problem!}.

Various approaches combining numerical and analytical methods
were proposed to derive Bell's inequalities with more than two
settings. In \cite{GISIN,MASSAR2} Bell's inequalities for two
parties and many measurement settings were derived. Recently Wu and Zong
\cite{WZ} derived an inequality for three parties, which
involves 4 local settings for two observers and 2 settings for the
third one. This inequality can be violated by the states (\ref{state})
for all values of $\alpha$ and thus is a more stringent condition on
local realism than the all standard Bell inequalities \cite{WZB,WW}. In Ref.
\cite{MAREK2} a multipartite "functional" Bell inequality for a continuous
range of settings of the apparatus at each site was derived. It is stronger
than the standard Bell inequalities for certain classes of states,
such as a family of bound entangled states \cite{MAREK-ADITI}, 
and the GHZ states \cite{MAREK2}.

Here we give an analytical method for deriving Bell's
inequalities for $N\!>\!2$ parties and many measurement
settings. The inequalities involving many measurement
settings are generalizations of the standard inequalities. That is, 
by reducing
the number of settings to two for each observer one recovers the full set of
the standard Bell inequalities \cite{WW,WZB}. The
inequalities reveal violations of local realism for wide classes of states, 
for which the standard Bell's inequalities fail in this task. 
A derivation of the necessary and sufficient condition for the violation of the
inequalities will be presented in a separate publication \cite{WORK}.

Before we give the method for many measurement settings, it is worthwhile to
recall the method of obtaining the full set of correlation function Bell's
inequalities with two measurement settings per observer. Consider $N$ observers
and allow each of them to choose between two dichotomic observables,
determined by some local parameters denoted by $\vec{n}_1$ and
$\vec{n}_2$. We choose such a notation for
brevity; of course, each observer can choose independently two
arbitrary directions. The assumption of local realism implies
existence of two numbers $A^j_1$ and $A^j_2$, each taking values
+1 or -1, which describe the predetermined \footnote{It is well known that any 
stochastic local realistic theory can be
formulated with an underlying deterministic one. Thus, here, we consider only
the deterministic theories.} result of a
measurement by the $j$-th observer of the observable defined by
$\vec{n}_1$ and $\vec{n}_2$, respectively.
The following algebraic identity holds for the predetermined
results for a single run of the experiment \cite{WZB}:
\begin{equation}
\sum_{s_1,...,s_N =\pm1} S(s_1,...,s_N) \prod_{j=1}^N [ A^j_1 +
s_j A^j_2]=\pm 2^N, \label{INEQ2}
\end{equation}
where $S(s_1,...,s_N)$ stands for an arbitrary "sign" function of the
summation indices $s_1,...,s_N=\pm1$, such that its values are
only $\pm1$, i.e. $S(s_1,...,s_N)=\pm1$.
For a specific run of the experiment one can introduce the product of the
local results  $\prod_{j=1}^N A^j_{k_j}$, with $k_j\!=\!1,2$. The correlation
function is the average over
many runs of the experiment $ E_{k_1,...,k_N}=\langle  \prod_{j=1}^N A^j_{k_j}
\rangle_{avg}$. After averaging the expression (\ref{INEQ2}) over
the ensemble of the runs of the experiment one obtains the
following set of Bell inequalities for local realistic correlation functions
\footnote{This set of inequalities is a sufficient and necessary condition
for the correlation functions entering them to have a local realistic model.} 
\begin{equation} 
|\hspace{-1mm} \sum_{s_1,...,s_N \atop = \pm 1} \hspace{-1mm}
S(s_1,...,s_N) \hspace{-1mm} \sum_{k_1,...,k_N \atop = 1,2} \hspace{-1mm}
s^{k_1-1}_1... s^{k_N-1}_N E_{k_1,...,k_N}| \leq 2^N. \nonumber
\end{equation} 
Since there are $2^{2^N}$ different
functions $S$, the above inequality represent a set of
$2^{2^N}$ Bell inequalities \cite{WW,WZB}.

Following the above ideas, and generalizing the trick introduced in \cite{WZ}, 
we show how to obtain Bell's inequalities involving many
measurement settings. We explain the method for the case of three
observers, and then show how to generalize it to an arbitrary number of
observers.

Suppose that the first two observers are allowed to choose between
four settings $\{1,2,3,4\}$, and the third one between two settings $\{1,2\}$.
We denote the family of inequalities that will be obtained as $4 \times 4 \times 2$ 
(this family contains the inequality of Wu and Zong \cite{WZ}). To avoid too many indices we introduce a new 
notation for the local
realistic values: $A_1,A_2,A_3$ and $A_4$ stand for the predetermined results for the
first observer under the local setting $1,2,3$ and $4$, respectively,
$B_1,B_2,B_3$ and $B_4$ are the similar values for the second observer, and
$C_1$ and $C_2$ are the predetermined values for the third observer (for the given
run).

The local realistic results for
the pair of settings $1$ and $2$ of Alice and Bob satisfy the identity 
(\ref{INEQ2}), i.e.
\begin{equation}
A_{12,12;S} \!\equiv \! \hspace{-2mm}\sum_{s_1,s_2=\pm 1}\!
S(s_1,s_2) (A_1+s_1A_2) (B_1+s_2B_2)=\pm 4, \label{kraj}
\end{equation}
where $1$ and $2$ are chosen from a larger set of four
measurement settings $\{1,2,3,4\}$. Similarly one 
defines $A_{34,34;S'}$ by replacing $A_1,A_2,B_1,B_2$ by $A_3,A_4,B_3,B_4$ 
respectively and $S$ by $S'$. Depending  on the value of $s= \pm 1$ one has
$(A_{12,12,S} + s ~ A_{34,34,S'})= \pm 8, ~~ \textrm{or} ~~ 0$. 
Therefore, the following algebraic identity holds:
\begin{equation}
\sum_{s'_1,s'_2 = \pm 1}S"(s'_1,s'_2)(A_{12,12,S}+s'_1
A_{34,34,S'})(C_1+s'_2C_2)=\pm 16. \label{gen}
\end{equation}
With the use of (\ref{gen}), after averaging over the runs, we can generate a family of
$(2^4)^3=2^{12}$ new Bell's inequalities \footnote{One has $2^4$ different
functions $S$ (and also $S'$ and $S''$), thus the full multiplicity is $2^{12}$}. 
They form a necessary condition for local realistic description to hold. Note
that inequalities involving three settings for a given observer can also be
obtained by, e.g., choosing the settings 2 and 3 identical.

Let us give an example. The Wu-Zong inequality is obtained if one chooses
$S"(1,1)=S"(1,-1)=S"(-1,1)=-S"(-1,-1)$. In such a case one obtains
\begin{eqnarray}
\hspace{-2mm} \sum_{s_1,s_2=\pm 1} \hspace{-5mm} & &  S(s_1, s_2)
(A_1 + s_1 A_2)
(B_1 + s_2 B_2) (C_1 + C_2)  \label{PPPM} \\
\hspace{-2mm}+\hspace{-2mm} \sum_{s_1,s_2=\pm 1} \hspace{-5mm} &
& S'(s_1,s_2) (A_3 + s_1 A_4) (B_3 + s_2 B_4) (C_1- C_2)\!=\!\pm
8. \nonumber
\end{eqnarray}
If one now puts $S' = \pm S''$ and $S = \pm S''$, and averages the
resulting algebraic identity over the runs, one obtains
\begin{eqnarray}&& |\sum_{g=1,2}(E_{11g} + E_{21g} + E_{12g}  -
E_{22g})| \label{glava} \\  
&+& |\sum_{g=1,2} (-1)^g(E_{33g} + E_{43g} +
E_{34g} - E_{44g})| \leq 4, \nonumber 
\end{eqnarray}
which is equivalent to the inequality of \cite{WZ}.

For $N=3$ Eq.(\ref{gen}) leads to the three families of
non-trivial Bell inequalities. One can generate the full set of members 
of a family by permuting the local settings in an inequality belonging 
to the given family. In the table we give representative inequalities of the 
families, and their maximal quantum values.
\begin{widetext}
\begin{table}[htbp]
\begin{tabular}{c c}
Typical inequality  & Max. quantum value \\
\hline
$ |-E_{111}-E_{331}+E_{112}-E_{332}| \leq 2$& $2\sqrt{2}$   \\
$ |-2E_{111}-E_{331}-E_{431}-E_{341}+E_{441}+2E_{112}-E_{332}-E_{432}-E_{342}+E_{442}| \leq 4$   & $4\sqrt{3}$   \\
$|-E_{111}-E_{211}-E_{121}+E_{221}-E_{331}-E_{431}-E_{341}+E_{441}$ & 8 \\
$+E_{112}+E_{212}  +E_{122}-E_{222}-E_{332}-E_{432}-E_{342}+E_{442}| \leq4$ &  
\end{tabular}
\end{table}
\end{widetext}
The first family of Bell's inequalities involves two settings for each observer
(we shall denote this property by $2 \times 2 \times 2$) 
and thus belongs to the standard
inequalities. Yet, the second and third family are new, of the type 
$3 \times 3 \times 2$ and $4 \times 4 \times 2$, respectively.

The results for some classes of 
states are summarized below. The new Bell inequalities often give more stringent
conditions on local realism than the standard inequalities. All numerical 
results are obtained using the ``amoeba" procedure.

(1) The maximal violations occur for the GHZ \cite{GHZ} states:
$|\psi\rangle \!=\! 1/\sqrt{2} (|000\rangle\!+\! |111\rangle)$
for all families.

(2) For $ \rho \!=\! (1\!-\!f)
|\psi\rangle \langle \psi|\! +\!  f \openone /8,$ where $|\psi\rangle$ is
the GHZ state, and $\openone/8$ is the completely
mixed state (noise), the highest possible fraction of noise such that the
state  still does not allow a local realistic description is
$f\!<\! \frac{1}{2}$ for both the standard and new Bell inequalities.

(3) Both the second and third family are violated
by the generalized GHZ states (\ref{state}) for the whole range
of $\alpha$. For $0 \le \alpha \le \frac{\pi}{12}$, that is, the range of the parameter, 
for which none of the standard Bell inequalities \cite{WZB,WW} is violated \cite{ZBLW},
the third family is violated by the factor of $\sqrt{1+\sin^2 2\alpha}$. 
This shows that in this case 
the family is a {\it stronger entanglement witness} than the full set
of $(2^{2^3}\!=\!256)$ standard Bell's inequalities \cite{WW,WZB}.

(4) The $W$ state: $|W\rangle\! =\!
1/\sqrt{3} (|001\rangle \!+ \!|010\rangle\! + \!|100\rangle)$ \cite{W}
violates the inequalities of the third family by the
factor of $1.7449$, whereas for the standard correlation function
Bell inequalities the factor is only $1.523$. The highest possible fraction of the white noise
that can be admixed to the $W$ state such that the resulting state still violates
the inequality is $f<0.4269$ for the new inequalities, whereas in the case of
standard inequalities $f<0.3434$.

(5) One way to show impossibility of local realistic description of mixed entangled 
states is to distill from them pure entanglement, that can 
violate Bell's inequality. Yet this is not possible for bound entangled states.
They are usually tested via various Bell's inequalities \cite{DUR}. We have
tested a three-qubit bound entangled state that has tripartite but no
bipartite entanglement, i.e. the entanglement across any split into two
parties is zero \cite{BENNETT}. The state is $\rho = 1/4 (\openone
\!- \!\sum_{j=1}^{4} |\psi_j\rangle \langle \psi_j|),$ where
$|\psi_1\rangle\!=\!|01+\rangle$, $|\psi_2\rangle\!=\!|1+0\rangle$,
$|\psi_3\rangle\!=\!|+01\rangle$ and $|\psi_4\rangle\!=\!|---\rangle$ with
$|\pm\rangle\!=\!(|0\rangle \pm |1\rangle)/\sqrt{2}$.  However, both the
standard and new Bell's inequalities are satisfied.

The method just applied can be
generalized in various ways to different numbers of measurement settings 
and to arbitrary number of observers. The following examples illustrate the strength of the
method.

(a) $3 \times 3 \times 2$-type inequalities: The inequalities
involving 3 settings for the first two observers and 2 settings
for the last one can be obtained by, e.g., choosing the
settings 1 and 2 identical for the two observers. If we put
$A_2\!=\!A_1$ and $B_2\!=\!B_1$ in Eq. (\ref{PPPM}), 
after averaging over the runs, 
we obtain the following inequality 
\begin{eqnarray*}
&& 4|\sum_{k=1,2} E_{11k}| \\ &&+ 
\sum_{s_1,s_2=\pm1}
|\sum_{i,j=3,4} \sum_{k=1,2}s_1^{i-1}s_2^{j-1}(-1)^{k-1}E_{ijk}| \le 8. \nonumber
\end{eqnarray*}
It is violated by the
 states (\ref{state}) for the full range of
$\alpha$. Again in the range of $0 < \alpha < \pi/12$, the violation is by the 
factor $\sqrt{1+\sin^2 2\alpha}$. Thus, in this case, the $3 \times 3
\times 2$ inequalities are just  as efficient as the $4 \times 4 \times 2$
ones.

(b) $4 \times 4 \times ... \times 4 \times 2$-type of inequality:
The multipartite inequalities involving 4 measurement settings for the first
$N - 1$ observers and 2 settings for the last observer can be derived as follows.
Analogously as in the case of two observers in Eq. (\ref{kraj}), for
the local realistic predetermined values for the $N\!-\!1$ observers
one can introduce
\begin{equation}
A_{12,...,12,S} = \sum_{s_1, ..., s_{N-1}=\pm 1} S(s_1, ..., s_{N-1})
\prod_{j=1}^{N-1} (A_j^1 + s_j A_j^2) \nonumber
\end{equation}
where $1$ and $2$ are two local settings chosen from a
larger set of four settings, $\{1,2,3,4\}$, for the $j$-th observer. 
By Eq.(\ref{INEQ2}) one has $A_{12,...,12,S} = \pm
2^{N-1}$. Similarly, the local realistic results for the pairs of
settings 3 and 4, and a different sign function $S'$, satisfy the
identity $A_{34,...,34,S'} = \pm 2^{N-1}$. One has $
[A_{12,...,12,S} + s ~ A_{34,...,34,S'}] \!=\! \pm 2^{N}$ or $0$,
depending on the value of $s= \pm 1$. By including the $N$-th
observer, who can choose between 2 measurement settings $\{1,2\}$ one 
 obtains 
\begin{eqnarray}
\sum_{s_1, s_2=\pm1} S(s_1,s_2) (A_{12,...,12,S} + s_1
A_{34,...,34,S'})\nonumber \\ \times (A_N^1 + s_2 A_N^2) = \pm 2^N.
\end{eqnarray} 
One can use this expression for generating new Bell inequalities
for $N$ observers in the same way as it was previously done in
Eq. (\ref{gen}-\ref{glava}) for three observers.

(c) $2^{N-1} \times 2^{N-1} \times 2^{N-2} \times 2^{N-3} \times
... \times 2$-type of inequality: The method can also be applied
to obtain Bell's inequalities involving exponential (in $N$) number of
measurement settings. The starting point is the identity (\ref{gen}) of the
type $4 \times 4 \times 2$ as derived above. 
One can introduce a similar identity for the settings $5$, $6$, $7$, $8$, for
the first  two observers, and $3$, $4$, for the third one. One can allow
the forth observer to choose  between two settings. Applying the same method as
the one described in (b),  one obtains an algebraic
identity which generates Bell's inequalities of the $8 \times 8 \times 4
\times 2$ type. One may apply the method {\it
iteratively}, increasing the number of observers by one, to obtain 
inequalities of the type given above.

It is clear that the method can be extended to various combinations of the
numbers of measurement settings, and observers. In Ref. \cite{LANL} one can
find another application of the method: a family of Bell inequalities for
$N\!=\!5$ qubits, which involves 8 settings for first two observers and, 4
settings, for the other three.

In summary, we present new Bell's inequalities
involving many measurement settings and prove that they give
more stringent conditions on the possibility of a local realistic
description of quantum states, than the standard Bell's inequalities
for two settings per observer.

Let us remark on some practical implications of our results
for communication complexity problems (problems of computing
a function if its inputs are distributed
among separated parties \cite{YAO}). In Ref. \cite{BZJZ} it was proven that
for every Bell's inequality there exists a communication
complexity problem, for which the protocol assisted by states
which violate the inequality is more efficient than any classical
one. The number of local measurement settings involved in the
inequality corresponds to the number of different values that one of the local inputs, on
which the function depends, may have. Since the generalized GHZ
states,  despite being pure and entangled, 
satisfy all standard Bell's inequalities, 
the quantum protocols utilizing them cannot have any advantage
over classical ones, as long as, the local inputs of the function have only
two possible values. However, a byproduct of the analysis given above
is that if one considers 3 or  more values for the local inputs,  a quantum
communication complexity protocols involving generalized GHZ states  can be
more efficient than any classical one (since the criterion of their
superiority is violation of the related Bell
inequality with 3 or more settings).

The work is supported by the Austrian-Polish project {\it Quantum
Communication and Quantum Information} (2002-2003). {\v C}.B. is
supported by the Austrian FWF project F1506, and by the European
Commission, Contract-No. IST-2001-38864 RAMBOQ. W.L. and T.P. are supported 
by the UG grant BW/5400-5-0256-3. M.\.Z. is supported by the Professorial Subsidy of FNP.


\end{document}